\documentclass[pra,twocolumn,amsmath,amssymb,superscriptaddress]{revtex4-1}

\usepackage{epsfig,amsmath}
\usepackage{subfigure}
\usepackage{graphicx}% Include figure files
\usepackage{dcolumn}% Align table columns on decimal point
\usepackage{stmaryrd}
\usepackage{mathrsfs}
\usepackage{pifont}
\usepackage{amsthm}
\usepackage{amssymb}
\usepackage{bm}
\usepackage{latexsym}
\usepackage[colorlinks=true,linkcolor=blue,citecolor=blue]{hyperref}
\usepackage{color}
\usepackage{epstopdf}

\begin{document}

\title{External-level assisted cooling by measurement}

\author{Jia-shun Yan}
\affiliation{Department of Physics, Zhejiang University, Hangzhou 310027, Zhejiang, China}

\author{Jun Jing}
\email{Email address: jingjun@zju.edu.cn}
\affiliation{Department of Physics, Zhejiang University, Hangzhou 310027, Zhejiang, China}

\date{\today}

\begin{abstract}
A quantum resonator in a thermal-equilibrium state with a high temperature has a large average population and is featured with significant occupation over Fock states with a high excitation number. The resonator could be cooled down via continuous measurements on the ground state of a coupled two-level system (qubit). We find, however, that the measurement-induced cooling might become inefficient in the high-temperature regime. Beyond the conventional strategy, we introduce strong driving between the excited state of the qubit and an external level. It can remarkably broaden the cooling range in regard to the non-vanishing populated Fock states of the resonator. Without any precooling procedure, our strategy allows a significant reduction of the populations over Fock states with a high excitation number, giving rise to nondeterministic ground-state cooling after a sequence of measurements. The driving-induced fast transition constrains the resonator and the ancillary qubit at their ground state upon measurement and then simulates the quantum Zeno effect. Our protocol is applied to cool down a high-temperature magnetic resonator. Additionally, it is generalized to a hybrid cooling protocol by interpolating the methods with and without strong driving, which can accelerate the cooling process and increase the overlap between the final state of the resonator and its ground state.
\end{abstract}

\maketitle

\section{Introduction}

Open quantum systems are prone to being thermalized due to their inevitable coupling with a finite-temperature environment~\cite{OpenQuantumSystem,QuantumNoise,QuantumOptics,CoolingLimit}. Pioneering experiments over the last decade have begun exploring the quantum behavior of microsize harmonic-resonator systems, adapting various techniques to cool specific modes far below the environmental temperature~\cite{NanomechanicalCooling,QuantumCoherentCoupling,LevitatedNanoparticleCooling}. Thus, the ground-state cooling and the more general pure-state preparation for a quantum system~\cite{DynamicalBackacionCooling,SidebandCoolingMechanicalMotion,TrappedIonLaserCooling,CollectiveTrappedIonLaserCooling} have long been a primary challenge and an indispensable task for the majority of quantum technologies. Cooling is a crucial requirement for the initialization of a quantum system, adiabatic quantum computing~\cite{RobustnessAdiabaticQC, AdiabaticEvolutionAlgorithm,OpenSystemsAdiabatic,EinsteinChannel,YouSuperconductingCircuits}, and ultrahigh-precision measurements~\cite{MeasurementWeakForceReviewBocko,MeasurementWeakForceReviewCaves}, to name a few.

Measurement-induced cooling~\cite{DemonLikeCooling,FeedbackCooling,QuantumMeasurementCooling,OneShotMeasurement,
CoolingQubit,PulsedOptomechanicsCooling} constitutes a class of subroutines among quantum control technologies based on measurement, including one-way computation~\cite{OneWayComputer,ExperimentalOneWayComputor}, the preparation of spatially localized states~\cite{SpatialCompression}, and entanglement generation~\cite{EntangleByCooling,EntangleByCooling2,EntanglementGeneration}. In a broader sense, these applications emerged from heated investigations~\cite{ZenoUnstableSystem,FromZenoToInverse,CriterionZenoAndInverse,PolarizationControlZeno} surrounding the quantum Zeno effect (QZE)~\cite{QuantumZeno,ThermodynamicControlZeno,ZenoParadox,ZenoeffectWineland}. Conventionally, the QZE states that measurements, such as the von Neumann projection and the generalized spectral decomposition~\cite{DynamicsZeno}, will affect in an essential way the dynamics of the measured system~\cite{TrappedIonMeasurementDynamics} and will hinder the decoherence process. Because cooling a system to a ground state is potentially connected to preparing an arbitrary pure state via a unitary operation, it is a direct consequence once the target system remains in its initial state because of QZE.

A typical and successful idea for cooling by measurement is to select the ground state of a continuous-variable target system, e.g., a mechanical resonator, out of an ensemble in a thermal state by carrying out continuous measurements on an ancillary system, e.g., a two-level system or qubit~\cite{PurificationViaMeasurements, MechanicalResonatorCooling,NonlinearMechanicalResonator,PulsedOptomechanicsCooling}. The whole system is forced to be in the ground state by the projective measurements with finite survival probability. This nondeterministic cooling strategy was theoretically proposed in Ref.~\cite{MechanicalResonatorCooling} and experimentally realized in Ref.~\cite{DemonLikeCooling}. Afterward, it was applied to cooling nonlinear mechanical resonators~\cite{NonlinearMechanicalResonator} and one-shot measurement cooling~\cite{OneShotMeasurement} to obtain the ground state. However, measurement-induced cooling assisted by a qubit might lose its efficiency when the initial temperature of the target system becomes too high. The occupations over the Fock states with a high excitation number might not be reduced, or the system will instead even be heated up by measurements.

In this work, we present a protocol allowing ground-state cooling of a harmonic resonator coupled to the transition of two levels in a three-level system. The state of the resonator-qubit subsystem can be protected by the strong driving between the excited state of the qubit and the extra or external level, which is decoupled from the target resonator. For the conventional repeated measurements over the ground state of the qubit, our protocol demonstrates a considerable advantage over the conventional one in regard to reducing the populations over the Fock states with a much greater excitation number. We practice our cooling protocol in the magnetic system quantized to be bosonic magnons. The magnon system is a significant interface that can be coupled with microwaves~\cite{PhotonicCavity,MicrowavePhotons,MicrowavePhotonsLimit,SpinPumping,CavityMagnomechanics}, electric currents~\cite{SpinPumping,TransmissionElectricalSignals,MagnonSpintronics}, mechanical motion~\cite{CavityMagnomechanics,UltrasonicWaves,ElasticExcitation,MagnonPolarons}, and light~\cite{BrillouinScattering,OpticalControlMagnetization,OpticalCoolingMagnons}. The varying magnetization induced by the magnon excitations inside the magnet, e.g., a yttrium iron garnet (YIG) sphere, leads to the deformation of its geometry structure, which forms the vibrational modes (phonons) of the sphere~\cite{UltrasonicWaves,MagnonEntanglement}. High temperature gives rise to (1) a large number of excitations and, consequently, a strong magnon-phonon interaction, making the mechanical degree of freedom non-negligible, and (2) a large linewidth~\cite{MicrowavePhotonsLimit}, reducing the coherent operation time. At a temperature on the order of 100~K, the average number of excitations in the magnon system could be hundreds. We find that after less than 100 equispaced unitary evolutions periodically interrupted by measurements of the ground state of the ancillary system, the effective temperature of the magnon mode could be reduced to several kelvins.

The rest of this work is arranged as follows. In Sec.~\ref{HamAndCoolOperater}, we introduce the model for a resonator coupled with a three-level system in the presence of a laser field resonantly driving the excited and external levels. The combination of the ground-state measurements of the ancillary system and the driving-induced resonator-qubit-state protection gives rise to a much broader cooling range in comparison to the conventional qubit-assisted protocol~\cite{MechanicalResonatorCooling}. The detailed derivation of the cooling coefficient is provided in the Appendix. In Sec.~\ref{CoolingMagnons}, we apply our cooling protocol assisted by the external level to the magnon system. Cooling performances by both the conventional and our protocol are compared with each other under various temperatures. Their distinction is also illustrated by the population histograms for the Fock states after a fixed number of measurements. In Sec.~\ref{DrivingStrength}, we propose a hybrid cooling protocol interpolating the conventional and our schemes. The cooling performance could then get promoted in regard to both speed and fidelity. The cooling mechanism of our cooling protocol is further analyzed in terms of a Zeno-like effect on the measured subspace under various driving strengths. In Sec.~\ref{Conclusion}, we discuss the distinction between the nondeterministic cooling by measurement and the sideband cooling in trapped-ion systems, stress the effect of the external driving, and then summarize the whole work.

\section{Cooling model and coefficient}\label{HamAndCoolOperater}

\begin{figure}[htbp]
\centering
\includegraphics[width=1\linewidth]{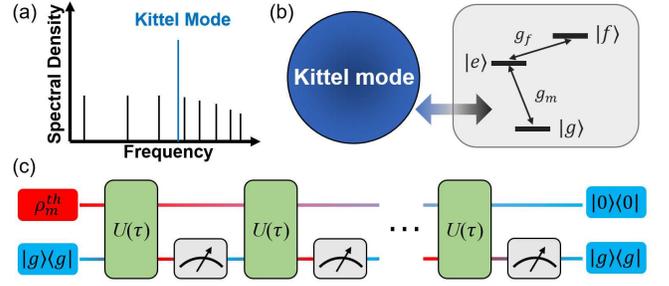}
\caption{(a) Sketch of the magnon spectrum. The dipole peak (Kittel mode) can be spectrally separated from the other magnon peaks under certain conditions~\cite{StrongSpinMagnonCoupling}. (b) Diagram of the model comprising a three-level system (the ancillary system) and a YIG sphere (the target system). $g_m$ is the coupling strength between them. The transition $|e\rangle\leftrightarrow|f\rangle$, driven by a laser field with a strength $g_f$, is far off resonant from the Kittel mode. (c) A general circuit model for the ground-state cooling by measurement. Initially, the resonator (the upper line) is in a thermal-equilibrium state, while the ancillary system (the bottom line) is at the ground state. Then the total system undergoes a free evolution for a period of $\tau$, after which an instantaneous projective measurement in the form of $|g\rangle\langle g|$ is performed. If the ancillary system is detected to remain at its ground-state, then the total system is allowed to evolve another period of $\tau$. The resonator is thus cooled down in a nondeterministic way and approaches its ground state $|0\rangle$ under the repetition of the preceding process.}\label{model}
\end{figure}

Our protocol [see Fig.~\ref{model}(c)] is an alternative realization of cooling a target resonator through the projective measurements of the ground state of the ancillary system, which is assisted by resonant driving between the excited state of the conventional ancillary qubit and an external state. In a realistic scenario, we consider a model consisting of a resonator [in the Kittel mode of Fig.~\ref{model}(a)] and an effective three-level system [a solid defect or emitter in Fig.~\ref{model}(b)]:
\begin{equation}\label{TotalHamiltonian}
\begin{aligned}
H&=H_0+H_I=\omega_mb^\dagger b +\omega_e|e\rangle\langle e| +\omega_f|f\rangle\langle f| \\
&+g_m(b+b^\dagger)\left(\sigma_{eg}^++\sigma_{eg}^-\right)+2g_f\cos{\upsilon t}\left(\sigma_{fe}^++\sigma_{fe}^-\right),
\end{aligned}
\end{equation}
where $b$ ($b^\dagger$) is the annihilation (creation) operator of the resonator with frequency $\omega_m$, $|e\rangle$ and $|f\rangle$ are the excited and external levels in the three-level system with frequencies $\omega_e$ and $\omega_f$, respectively, and are resonantly driven by a laser field with strength $g_f$ and frequency $\upsilon=\omega_f-\omega_e$. The energy of the ground state $|g\rangle$ of the ancillary system is assumed to be $\omega_g\equiv0$. The resonator is coupled to the transition $|e\rangle\leftrightarrow|g\rangle$ with a coupling strength $g_m$ and is far off resonant from the transition $|e\rangle\leftrightarrow|f\rangle$. Note the external level $|f\rangle$ is effectively decoupled from the resonator and both $|g\rangle$ and $|e\rangle$, which is not necessarily the highest level in the three-level system. The atomic transition operators are denoted by $\sigma^+_{ij}=|i\rangle\langle j|$ and $\sigma^-_{ij}=|j\rangle\langle i|$. Under $g_f=0$ and the rotating-wave approximation (RWA), the total Hamiltonian in Eq.~(\ref{TotalHamiltonian}) is reduced to the Jaynes-Cummings (JC) model with $H=\omega_e|e\rangle\langle e|+\omega_mb^\dagger b+g_m(\sigma_{eg}^+b+\sigma_{eg}^-b^\dagger)$, which is used in the conventional measurement-induced cooling scheme~\cite{MechanicalResonatorCooling}. In the hybrid magnon systems~\cite{StrongSpinMagnonCoupling,MagnonQubitcoupling} we are interested, the coupling strength $g_m$ is about $1-10$ MHz, which is much smaller than the frequencies of the magnetic resonator and qubit, $\omega_e, \omega_m\sim 10$ GHz. The RWA is thus justified in our model.

In the rotating frame with respect to $H_0=\omega_mb^\dagger b+\omega_e|e\rangle\langle e|+\omega_f|f\rangle\langle f|$ and under the RWA, the overall Hamiltonian reads,
\begin{equation}\label{rotatedHamiltonian}
\begin{aligned}
H'_I&=e^{iH_0t}He^{-iH_0t}-H_0 \\
&\approx g_m\left(\sigma_{eg}^+b+\sigma_{eg}^-b^\dagger\right)+g_f\left(\sigma_{fe}^++\sigma_{fe}^-\right),
\end{aligned}
\end{equation}
where we have applied the resonant condition $\omega_m=\omega_e$. With the Hamiltonian $H'_I$ in Eq.~(\ref{rotatedHamiltonian}) connecting the target system and the ancillary system, our protocol follows the general circuit model of the measurement-induced cooling. As shown in Fig.~\ref{model}(c), it is a concatenation of the free unitary evolutions under the effective Hamiltonian $H'_I$ and the repetition of the instantaneous measurements of the ground state of the ancillary system $M_g\equiv|g\rangle\langle g|$. Initially, the resonator [represented by the upper line in Fig.~\ref{model}(c)] is prepared in a thermal-equilibrium state, and the three-level system is in the ground state; then the overall density matrix reads
\begin{equation}\label{InitialState}
	\rho(0)=|g\rangle\langle g|\otimes\rho_m^{\rm th}.
\end{equation}
After each desired evolution period $\tau$, a projective measurement $M_g$ is performed on the discrete system to check whether it is still in the ground state. If so, the overall state takes the form
\begin{equation}
	\rho(\tau)=\frac{M_gU(\tau)\rho(0)U^\dagger(\tau)M_g}{{\rm Tr}[M_gU(\tau)\rho(0)U^\dagger(\tau)M_g]},
\end{equation}
where $U(\tau)=\exp(-iH'_I\tau)$ and the denominator is the probability of the ancillary system being its ground state after a single measurement. The ground-state cooling is essentially incoherent and probabilistic due to the uncertainty of the measurement results. After $N$ measurements, the total time taken by the cooling protocol is $t=N\tau$ under the assumption that the measurements are instantaneously accomplished. The success probability of detecting the ancillary system in its ground state $|g\rangle$ at time $t$ reads,
\begin{equation}\label{probability}
P_g(N)={\rm Tr}\left[V_g(\tau)^N\rho_m^{\rm th}V_g^\dagger(\tau)^N\right],
\end{equation}
where the nonunitary operator $V_g(\tau)\equiv\langle g|U(\tau)|g\rangle$ acts only in the resonator space. For the resonator prepared in the thermal state $\rho^{\rm th}_m=\sum_np_n|n\rangle\langle n|$ with $p_n=\bar{n}_{\rm th}^n/(1+\bar{n}_{\rm th})^{n+1}$ and $\bar{n}_{\rm th}\equiv{\rm Tr}(b^\dagger b\rho^{\rm th}_m)$, the probability of finding the resonator in its ground state is upper bounded by $p_0=1/(1+\bar n_{\rm th})$.

The evolution operator $V_g(\tau)$ is diagonal in the Fock-state basis $\{|n\rangle\}$:
\begin{equation}\label{Vg}
V_g(\tau)=\sum_{n\geq 0}\alpha_n(\tau)|n\rangle\langle n|,
\end{equation}
where $\alpha_n(\tau)$'s are defined as the cooling coefficients:
\begin{equation}\label{gfCoefficient}
\alpha_n(\tau)=\frac{\Omega_n^2+ng_m^2(\cos{\Omega_n\tau}-1)}{\Omega_n^2}, \quad \Omega_n\equiv\sqrt{g_f^2+ng_m^2}.
\end{equation}
And then $p_n\rightarrow|\alpha_n|^{2N}p_n$. More details can be found in the Appendix. In the absence of the driving field, i.e., $g_f=0$, the cooling coefficient will be reduced to that for the conventional cooling protocol~\cite{MechanicalResonatorCooling},
\begin{equation}\label{Coefficient}
\beta_n(\tau)=\cos\left(g_m\sqrt{n}\tau\right)
\end{equation}
for $\omega_m=\omega_e$. Note that $\alpha_0(\tau)=\beta_0(\tau)=1$, and $|\alpha_{n>0}(\tau)|$, $|\beta_{n>0}(\tau)|\leq1$ and both of them are independent of the initial temperature or the initial population distribution. For the current protocol, $|\alpha_n(\tau)|^2=1$ occurs with special $n$'s satisfying $\Omega_n\tau=2k\pi$, where $k$ is an arbitrary integer. However, for the conventional protocol, $|\beta_n(\tau)|^2=1$ when $g_m\sqrt{n}\tau=k'\pi$. It is therefore important to understand that both protocols have a finite-width applicable range. The width could be roughly regarded as the (quasi)period of those special $n$'s.

\begin{figure}[htbp]
\centering
\includegraphics[width=0.9\linewidth]{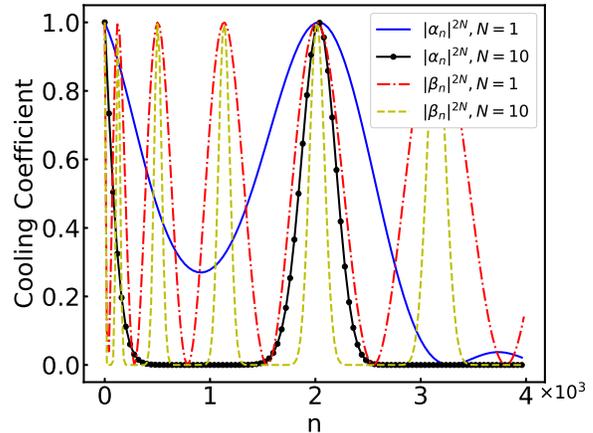}
\caption{Cooling coefficients as functions of the Fock-state index $n$ by a single measurement and $10$ measurements on the ground state of the ancillary systems. The blue solid line and the black-solid line with circles indicate $|\alpha_n(\tau)|^{2N}$ in our protocol using an ancillary three-level system. The red dash-dotted line and the yellow dashed line indicate $|\beta_n(\tau)|^{2N}$ in the conventional protocol using an ancillary two-level system. The coupling between the resonator and the ancillary system is $g_m/\omega_m =0.0004$, the driving strength is $g_f/g_m=50$, and the measurement period is $\omega_m\tau=700$.}	\label{CoolingCoefficient}
\end{figure}

The cooling range of $n$ is exclusively determined by $|\alpha_{n>0}(\tau)|^2=1$. According to Eq.~(\ref{gfCoefficient}) with $\Omega_n=2j\pi/\tau$, $j\in\mathbb{N}$, the population of the states $|n_j=[(2j\pi/\tau)^2-g_f^2]/g_m^2\rangle$ will survive under the measurement-induced cooling. These $j$'s for $n_j>0$ constitute a cooling-free set $A=\{j|j\in\mathbb{N}, j\geq g_f\tau/2\pi\}$. The other Fock states will be depopulated by measurements. In parallel, under the conventional cooling protocol, the protected Fock-states $|n_k=k^2\pi^2/(g_m^2\tau^2)\rangle$ constitute another set $B=\{k|k\in\mathbb{N}\}$. The quasi-periods for both protocols are then
\begin{equation}\label{Delta}
\Delta_{\alpha}=\frac{4(2j+1)\pi^2}{g_m^2\tau^2}, \quad \Delta_{\beta}=\frac{(2k+1)\pi^2}{g_m^2\tau^2}.
\end{equation}
By comparing $\Delta_{\alpha}$ and $\Delta_{\beta}$ and taking into account the starting elements in $A$ and $B$, $j\geq g_f\tau/2\pi$ and $k\geq1$, one can find strong driving $g_f$ yields a much wider cooling range than the conventional protocol.

In Fig.~\ref{CoolingCoefficient}, we plot $|\alpha_n|^{2N}$ and $|\beta_n|^{2N}$ for each $n$ with $N=1$ and $N=10$, respectively. For the ground state $|n=0\rangle$, both coefficients are 1, which is the foundation for all protocols of measurement-induced cooling. We focus on the first quasi-period bounded by the ground state and the first $|n\neq0\rangle$ at which $|\alpha_n|^2=1$ or $|\beta_n|^2=1$. With the parameters from the magnon system~\cite{StrongSpinMagnonCoupling}, the first nonvanishing solution under the conventional cooling protocol is $n\approx125$, and that under our protocol is about $n\approx2000$. Additionally, when $N=10$, $|\alpha_n|^{2N}<10^{-3}$ applies to the Fock states with $500<n<1350$. This means that the population of the resonator in this range can be significantly suppressed by only $10$ measurements. So the external-level assisted protocol outperforms the conventional protocol in the high-temperature regime.

In the large-$N$ limit, the nonunitary evolution operator becomes sparser and sparser,
\begin{equation}\label{VN}
V_g(\tau)^N=\sum_{n\geq 0}\alpha_n^N|n\rangle\langle n|\stackrel{N\rightarrow \infty}{\longrightarrow} |0\rangle\langle 0|+\sum_{j\in A}|n_j\rangle\langle n_j|.
\end{equation}
Consequently, the density matrix of the resonator becomes
\begin{equation}\label{DensityMatrix}
\begin{aligned}
\rho_m(N\tau)&=\left[V_g(\tau)^N\rho_m^{\rm th}V_g^\dagger(\tau)^N\right]/P_g(N)\\
&=\frac{\sum_{n\geq 0}|\alpha_n|^{2N}p_n|n\rangle\langle n|}{P_g(N)} \\
&\stackrel{N\rightarrow \infty}{\longrightarrow}\frac{p_0|0\rangle\langle 0|+\sum_{j\in A}p_{n_j}|n_j\rangle\langle n_j|}{P_L},
\end{aligned}
\end{equation}
where $p_{n\geq0}$ describes the initial distribution of the resonator over Fock states and the survival probability $P_g(N)$ of the resonator at $\rho_m(N\tau)$ is
\begin{equation}\label{Probability}
P_g(N)=\sum_{n\geq 0}|\alpha_n|^{2N}p_n\stackrel{N\rightarrow \infty}{\longrightarrow}p_0+\sum_{j\in A}p_{n_j}\equiv P_L.
\end{equation}
Under continuous measurements, the average population of the resonator becomes
\begin{equation}\label{Number}
\begin{aligned}
\bar{n}(N) &\equiv{\rm Tr}\left[b^\dagger b\rho_m(N\tau)\right]=\frac{\sum_{n\geq 0}n|\alpha_n|^{2N}p_n}{\sum_{n\geq 0}|\alpha_n|^{2N}p_n} \\
&\stackrel{N\rightarrow \infty}{\longrightarrow}\frac{\sum_{j\in A}n_jp_{n_j}}{P_L}=\frac{\sum_{j\in A}n_jp_{n_j}}{p_0+\sum_{j\in A}p_{n_j}}.
\end{aligned}
\end{equation}
This indicates that the final $\bar{n}(N)$ is directly determined by the cooling-free set of Fock states. As for the conventional cooling protocol, $j\in A$ in Eq.~(\ref{Number}) should be replaced by $j\in B$. For a comparatively low temperature with limited $\sum_{j\in A}p_{n_j}$, $\bar{n}(N\to\infty)\approx0$ can always be achieved by either cooling protocol. According to Fig.~\ref{CoolingCoefficient} and Eq.~(\ref{Number}), however, the ground state cooling will become difficult or even impossible in the presence of the non-negligible populations $p_{n_j}$ over the cooling-free Fock states. With a significantly broadened cooling range covering more Fock states with a higher excitation number, it is shown in the next section that our external-level assisted protocol can realize the ground state cooling in a much higher temperature regime. The magnon system is chosen to be the target resonator.

\section{Measurement cooling of magnons}\label{CoolingMagnons}

\begin{figure}[htbp]
\centering
\includegraphics[width=0.8\linewidth]{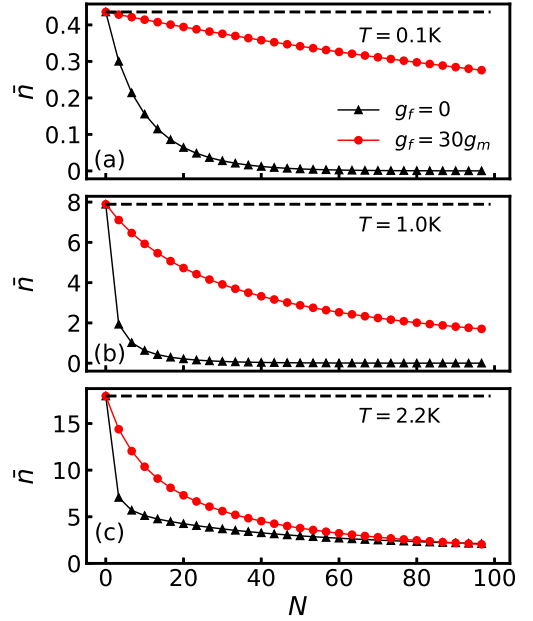}
\caption{The average magnon number $\bar{n}(N)$ as a function of measurement numbers $N$. The black solid lines marked with triangles and the red solids line marked with circles indicate the conventional and external-level assisted cooling protocols, respectively. (a) $T=0.1$ K, (b) $T=1.0$ K, and (c) $T=2.2$ K. The frequency of the Kittel mode is set as $\omega_m=15.6$ GHz. $g_m=2\pi\times1$ MHz, $g_f/g_m=30$, and $\omega_m\tau=700$. }	\label{Cooling_low_temp}
\end{figure}

The general nondeterministic protocol described by the nonunitary operator $V_g$ in Eq.~(\ref{Vg}) and its cooling coefficient $\alpha_n$ in Eq.~(\ref{gfCoefficient}) is now applied to the magnon mode. Normally, the magnon system is set up by a sphere made of a ferrimagnetic insulator YIG, which has a giant magnetic quality factor ($\sim10^5$) and supports ferromagnetic magnons with a long coherence time ($\sim 1 \mu$s)~\cite{MagnonQuanlity,Magnonics,MagnonDarkModes}. The paradigmatic model described by Eq.~(\ref{TotalHamiltonian}) is then physically relevant to a magnetic emitter (the ancillary system) coupled to the Kittel mode (the target resonator) via the induced magnetic fields by putting a diamond nitrogen-vacancy center near a YIG sphere~\cite{StrongSpinMagnonCoupling}. The Kittel mode, in which all the spins precess in phase and with the same amplitude, can be efficiently excited by an external magnetic field and becomes separable from the other modes in the spectrum due to its dipolar character, as shown in Fig.~\ref{model}(a). If the excited state of the three-level system is resonant with the Kittel mode ($\sim 10$ GHz), then the coupling strength could approach a strong-coupling regime ($g_m\approx 10$ MHz, much larger than the decay rate $\Gamma\approx0.2$ MHz)~\cite{StrongSpinMagnonCoupling}. A laser field, whose Rabi frequency is much stronger than the coupling strength between the magnetic resonator and the three-level system, is assumed to be resonant with the transition $|e\rangle\leftrightarrow|f\rangle$ and far off-resonant from $|g\rangle\leftrightarrow|e\rangle$, as shown in Fig.~\ref{model}(b).

In Fig.~\ref{Cooling_low_temp}, we first check the cooling efficiency with the average population $\bar{n}(N)$ of the magnon resonator under the conventional and external-level assisted protocols. The resonator is prepared in a low-temperature regime. In Fig.~\ref{Cooling_low_temp}(a) with $T=0.1$ K, the conventional protocol with $g_f=0$ demonstrates a significant advantage over our protocol with $g_f/g_m=30$. It is found that after $N=60$ measurements, $\bar{n}$ is reduced from $0.44$ to $0.002$ by the former protocol, and in contrast, it is only reduced to merely $0.33$ by the latter one. This result can be expected from the cooling-coefficient distribution in Fig.~\ref{CoolingCoefficient}. The conventional protocol is featured with a steeper slope in a limited range of Fock states, indicating a stronger cooling efficiency in the space with a small excitation number, i.e., in the low-temperature regime. The cooling efficiencies of these two protocols are found to move gradually closer to each other with an increasing initial temperature of the target magnon. In Fig.~\ref{Cooling_low_temp}(b), with $T=1.0$ K, $\bar{n}$ can be reduced to about $0.003\%$ and $0.3\%$ of its initial value after $N=80$ measurements by the conventional and external-level assisted protocols, respectively. However, in Fig.~\ref{Cooling_low_temp}(c), with $T=2.2$ K, the average populations $\bar{n}(N)$ under both protocols converge after $80$ measurements. For a typical hybrid system of a magnon coupled with an emitter, it is numerically found that both measurement-cooling protocols exhibit almost the same cooling efficiency at about a few kelvins, which could be regarded as a critical temperature.

\begin{figure}[htbp]
\centering
\includegraphics[width=0.8\linewidth]{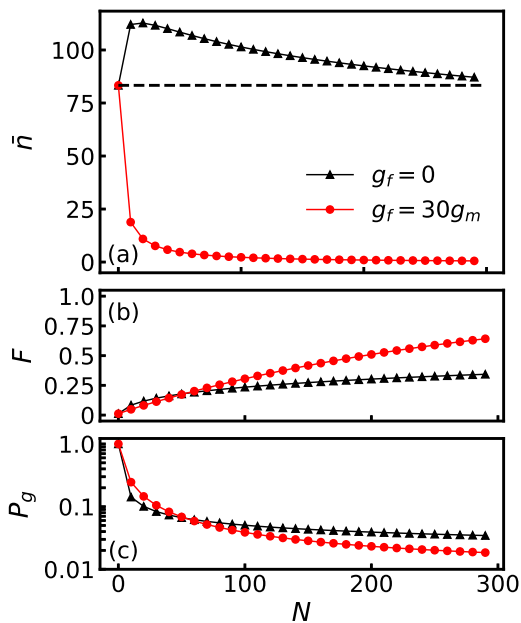}
\caption{Cooling performances at a higher temperature $T=10$ K of both measurement-cooling protocols, including (a) the average magnon number $\bar{n}$, (b) the fidelity of the ground state $F(N)\equiv\langle0|\rho_m(N\tau)|0\rangle$, and (c) the survival probability $P_g(N)$ of detecting the ancillary system at $|g\rangle$ and the resonator in $\rho_m(N\tau)$ as functions of $N$. $\omega_m=15.6$ GHz, $g_m=2\pi\times1$ MHz, $g_f/g_m=30$, and $\omega_m\tau=700$. }\label{Cooling_high_temp}
\end{figure}

\begin{figure}[htbp]
\centering
\includegraphics[width=0.8\linewidth]{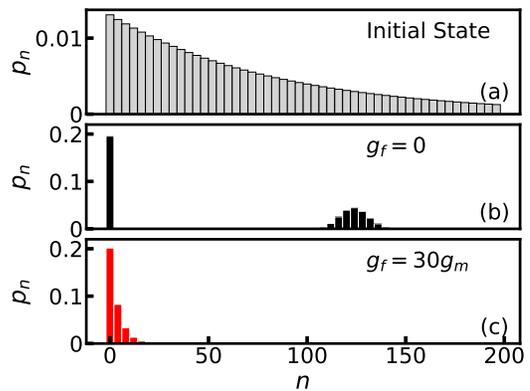}
\caption{Population histograms for various states of the resonator over the Fock space. (a)The initial thermal state at $T=10$ K. (b) and (c) The states after $N=60$ measurements by the conventional protocol and our protocol, respectively. $\omega_m=15.6$ GHz, $g_m=2\pi\times1$ MHz, and $\omega_m\tau=700$. }	\label{Distribution}
\end{figure}

The advantage of our external-level-assisted protocol manifests in a high-temperature regime as indicated by the dramatically expanded cooling range in Fig.~\ref{CoolingCoefficient}. Figure~\ref{Cooling_high_temp}(a), with $T=10$ K, demonstrates a completely inverted pattern of Fig.~\ref{Cooling_low_temp}(a), with $T=0.1$ K. Note the other parameters remain invariant. In Fig.~\ref{Cooling_high_temp}(a), the conventional cooling strategy is found to heat up rather than cool down the resonator. The average population $\bar{n}$ firstly increases to about $113$ during the first $20$ measurements and then gradually decreases to its initial value $\bar{n}\approx83$ after $300$ measurements. In sharp contrast, our cooling protocol assisted by a driven three-level system keeps reducing the average population to $\bar{n}\approx4$ after $60$ measurements, and down to $\bar{n}\approx0.5$ after $300$ measurements. The advantage of our external-level-assisted protocol is also shown by the ground state fidelity. In Fig.~\ref{Cooling_high_temp}(b), the fidelity of our protocol keeps increasing to $0.7$ after $300$ measurements; while in the same time it is less than $0.4$ under the conventional protocol. Yet both nondeterministic protocols are found to be source exhaustive in terms of the low survival or success probability of measurements $P_g(N)$. In Fig.~\ref{Cooling_high_temp}(c), both $P_g(N)$'s decrease to less than $0.1$ after $40$ measurements. And after $50$ measurements, the survival probability of our protocol, i.e., the probability of cooling the resonator to its ground state, becomes smaller than the conventional one.

The remarkable distinction between the conventional cooling method and our method in the high-temperature regime can also be illustrated by population histograms for various states of the resonator in the Fock space. As shown in Fig.~\ref{Distribution}(a), the resonator of the initial state is widely distributed, which has a considerable population even at $|n=200\rangle$. After $N=60$ measurements, one can see in Fig.~\ref{Distribution}(b) that nearly $20\%$ of the population aggregates at the ground state $|n=0\rangle$ and the rest are centered around $|n=125\rangle$. Apparently, Fock states with such high excitation numbers are cooling free in the conventional protocol. However in Fig.~\ref{Distribution}(c), nearly $60\%$ of the population aggregates around the ground state in our protocol, demonstrating a clear cooling effect in a wide range of $n$. Over the state $|n=14\rangle$, the population becomes less than $0.01$.

Moreover, one can find that the last state is still a thermal state. Using the parameters in Fig.~\ref{Distribution}, it is found that $\cos(\Omega_n\tau)\ll1$ and $ng_m^2/\Omega_n^2<ng_m^2/(g_f^2+g_m^2)\ll1$ for $n\leq20$. Then according to Eq.~(\ref{gfCoefficient}), the population of the resonator at the Fock state $|n\rangle$ becomes
\begin{equation}
\begin{aligned}
p_n\rightarrow&|\alpha_n|^{2N}p_n\propto \left(1-ng_m^2/\Omega_n^2\right)^{2N}\exp\left(-\frac{n\hbar\omega_m}{k_BT}\right)\\
&\approx \exp\left(-\frac{2ng_m^2N}{g_f^2+g_m^2}\right)\exp\left(-\frac{n\hbar\omega_m}{k_BT}\right) \\ &=\exp\left\{-\frac{n\hbar\omega_m}{k_B}\left[\frac{1}{T}+\frac{2k_Bg_m^2N}{\hbar\omega_m(g_f^2+g_m^2)}\right]\right\}\\
&=\exp\left(-\frac{n\hbar\omega_m}{k_BT_{\rm eff}}\right)
\end{aligned}
\end{equation}
up to a normalization coefficient. So $p_n$ still follows a Maxwell-Boltzmann distribution, and the resonator can be seen at an effective temperature $T_{\rm eff}$, which can also be defined by the average population $\bar{n}$ as
\begin{equation}\label{Teff}
T_{\rm eff}=\frac{\hbar\omega_m}{k_B\ln(1+1/\bar{n})}.
\end{equation}
Then one can estimate the overlap of any cooled state and the thermal state with $T_{\rm eff}$ by the fidelity
\begin{equation}\label{Fth}
F_{\rm th}\equiv\left\{{\rm Tr}\left[\sqrt{\sqrt{\rho_m(N\tau)}\rho_m^{\rm th}(T_{\rm eff})\sqrt{\rho_m(N\tau)}}\right]\right\}^2.
\end{equation}
By virtue of Eqs.~(\ref{Teff}) and (\ref{Fth}), we find the effective temperature in Fig.~\ref{Distribution}(c) is about $0.6$ K with a near-unit fidelity.

\begin{figure}[htbp]
\centering
\includegraphics[width=0.9\linewidth]{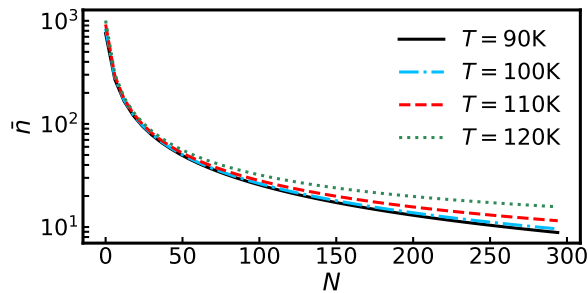}
\caption{Average magnon numbers $\bar{n}(N)$ under various initial high temperatures. The black solid, blue dot-dashed, red dashed, and green dotted lines describe $T=90$, $100$, $110$ and $120$ K, respectively. $\omega_m=15.6$ GHz, $g_m=2\pi\times1$ MHz, $g_f/g_m=50$, and $\omega_m\tau=220$.}\label{Diff_temp}
\end{figure}

In Fig.~\ref{Diff_temp} for the resonator in an even higher temperature regime, a decent cooling performance of our protocol can be observed at a temperature around 100 K, although it decreases slightly with increasing $T$. When $T=100$ K (see the blue dot-dashed line), the average population for such a resonator with a frequency on the order of $10$ GHz could be reduced from $\bar{n}\approx800$ to $\bar{n}\approx50$ by $50$ measurements and continuously down to $\bar{n}\approx10$ by $300$ measurements. According to the effective temperature defined in Eq.~(\ref{Teff}), these two values of $\bar{n}$ correspond to $T_{\rm eff}=6.02$ K and $T_{\rm eff}=1.25$ K, respectively. Thus, for a magnon resonator prepared at the same order of room temperature, our external-level assisted strategy is a promising choice to approach ground-state cooling without a refrigerator, although at a cost of continuous measurements.

\section{Hybrid cooling protocol and subspace protection by driving}\label{DrivingStrength}

Due to the distinct cooling ranges, the conventional and current protocols are found to exhibit a cooling advantage in the low- and high-temperature regimes, respectively. In this section, we propose a hybrid cooling scheme by interpolating the methods with and without driving between the excited level $|e\rangle$ and the external level $|f\rangle$. It accelerates the cooling process by reducing the desired numbers of measurements $N$ to approach ground-state cooling.

\begin{figure}[htbp]
\centering
\includegraphics[width=0.8\linewidth]{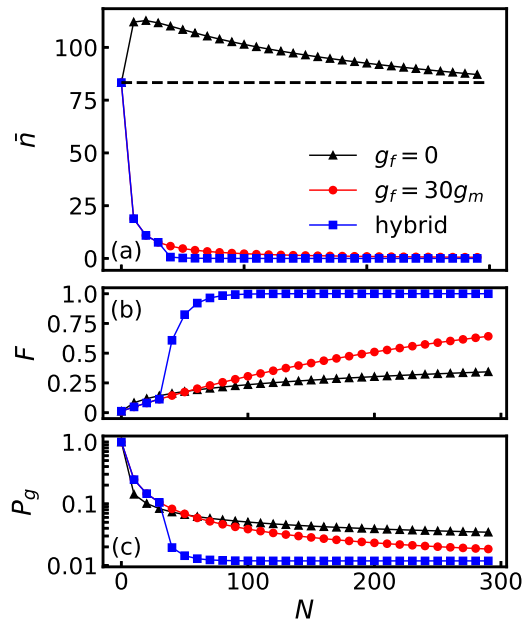}
\caption{Cooling performance at $T=10$ K, including (a) the average magnon number $\bar{n}$ (b) the ground state fidelity $F$, and (c) the survival probability $P_g$ as functions of $N$, under various protocols. The black lines marked with triangles, the red lines marked with circles and the blue lines marked with squares denote the conventional, external-level-assisted, and hybrid cooling protocols, respectively. $\omega_m=15.6$ GHz, $g_m=2\pi\times1$ MHz, and $\omega_m\tau=700$. }	\label{mix}
\end{figure}

\begin{figure}[htbp]
\centering
\includegraphics[width=0.8\linewidth]{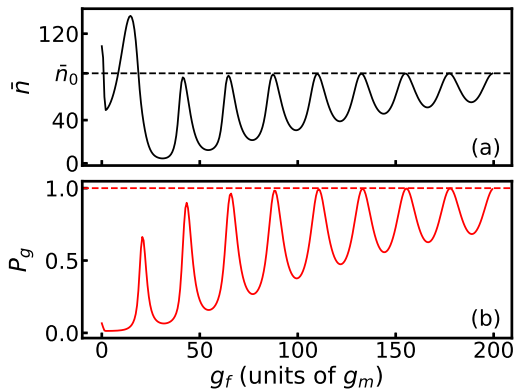}
\caption{(a) The average magnon number $\bar{n}$ and (b) the survival probability $P_g$ as functions of the normalized driving strength $g_f$ (in units of $g_m$). The number of measurements is fixed at $N=50$. $\omega_m=15.6$ GHz, $g_m=2\pi\times1$ MHz, and $\omega_m\tau=700$. }\label{strong_driving_limit}
\end{figure}

In particular, to cool down a resonator initially prepared at a comparatively higher temperature, e.g., $T=10$ K, one can first employ the strategy with driving, taking advantage of its wide cooling range; and then turn off the driving laser at a desired interpolating point and continue to cool the resonator using the conventional strategy without driving, which is more favorable in a low-temperature regime. In Fig.~\ref{mix}, the hybrid protocol is shown under the same conditions as in Fig.~\ref{Cooling_high_temp}. When the average magnon number declines to $\bar{n}=10$ (about one order smaller than its initial value) by $N=30$ measurements, the driving is turned off, i.e., $g_f=0$. We then find that $\bar{n}$ [see Fig.~\ref{mix}(a)] decreases with a remarkably faster rate than before. And in the same time, the fidelity of the resonator in its ground state $|0\rangle$ [see Fig.~\ref{mix}(b)] approaches 1 in a sharper way. Although the survival probability becomes worse [see Fig.~\ref{mix}(c)], which is about $0.01$, the hybrid protocol still exhibits a better cooling performance than both the conventional and external-level assisted protocols. It is capable of reducing the average magnon number from $83$ to $0.09$, i.e., almost $4$ orders in magnitude, with a high fidelity $F\approx0.92$ through only $N=60$ measurements. Therefore, the driving protocol we proposed can be used as a pre-cooling procedure for the conventional one.

We now scrutinize the effect of the driving strength $g_f$ in addition to broadening the cooling range. When the driving strength $g_f$ between the excited state $|e\rangle$ and the external state $|f\rangle$ overwhelms the coupling strength $g_m$ between the resonator and the ancillary system, the fast transition $|e\rangle\leftrightarrow|f\rangle$ suppresses the state leakage out of the subspace consisting of the resonator and level of $|g\rangle$ the three-level system~\cite{ActingOutside}. On repeated measurements by $M_g$, the dynamics of the nondeterministic state $|g\rangle\langle g|\otimes\rho_m(N\tau)$ will be consequently inhibited by the driving. Thus, one can see in Fig.~\ref{strong_driving_limit}(b) that the survival probability $P_g$ of detecting the ground state $|g\rangle$ is asymptotically enhanced by increasing $g_f/g_m$. For the resonator part, its average population $\bar{n}$ in Fig.~\ref{strong_driving_limit}(a) follows roughly the same pattern as $P_g$, where the black dashed line denotes the initial $\bar{n}$ for $T=10$ K. It is interesting to see that (1) when $g_f/g_m<20$, the cooling protocol might actually heat up the target resonator, meaning cooling demands a sufficiently strong driving, and (2) $\bar{n}$ does not follow a monotonic relation with $g_f/g_m$ under a fixed $N$, meaning a larger $g_f/g_m$ is not necessarily more favorable than a smaller $g_f/g_m$ in cooling efficiency.

\section{Discussion and conclusion}\label{Conclusion}

The underlying mechanism in our context of measurement-induced cooling or in the context of measurement-induced purification~\cite{PurificationViaMeasurements} is nondeterministic. Each measurement after an interval $\tau$ of free unitary evolution performs a postselection on the whole system, by which the distribution of the target resonator over Fock states with a high excitation number is discarded and only the ground state is collected from the ensemble. The success of detecting the system in the ground state cannot be guaranteed for every measurement. The limited survival probability $P_g(N)$ in Eq.~(\ref{probability}) and the resulting nonunitary evolution operator $V_g(\tau)$ in Eq.~(\ref{Vg}) indicate the cost for all similar protocols. They are profoundly distinct from the laser-cooling schemes used in trapped-ion systems~\cite{TrappedIonLaserCooling,CollectiveTrappedIonLaserCooling}. Motional cooling can be achieved by the imbalanced transition of the resonator Fock states due to the existence of a spontaneous decay channel. In this work, however, measurement by $M_g$ forces the resonator to the ground state in a nondeterministic way~\cite{Postselection,MechanicalResonatorCooling}.

The conventional measurement-induced cooling becomes inefficient when the temperature becomes higher than a critical value due to the limited cooling range. We found that the cooling-by-measurement effect on the ground state $|0\rangle_m|g\rangle$ in the subspace consisting of the resonator and the ground level of the ancillary system can be sustained or strengthened by strongly driving the excited level $|e\rangle$ of the ancillary system and an external level $|f\rangle$. Intuitively, once $|e\rangle$ is occasionally populated by the unwanted transition via the interaction with the resonator, the extra energy is rapidly pumped out of the subspace. Formally, the strong driving serves as an extra measurement of the ground state of the subspace. Then the cooperation of the existing projective measurement and the driving leads to the extension of the cooling range to the high-temperature regime around $100$ K.

In conclusion, we provided an external-level assisted protocol realizing the ground state cooling of a thermal resonator. It significantly outperforms conventional cooling by measurement in the high-temperature regime. As an application, the magnon resonator in the Kittel mode can be effectively cooled down when $T$ is over $10$ K and even $100$ K using our protocol. In addition, a hybrid scheme was proposed to demonstrate an even better cooling performance in both the average excitation number and the ground state fidelity. Our work essentially provides a general and efficient pre-cooling procedure for the conventional measurement-induced cooling protocols, as an alternative scheme without using an expensive dilution refrigerator. The application range of our protocol for a surrounding temperature on the order of $100$ K might indicate a possible explanation for the quantum effect in some biological systems~\cite{QuantumBiology}. It also justifies that a continuous strong driving can simulate a Zeno-like effect or projective measurement on a pure state.

\section*{Acknowledgments}

We acknowledge grant support from the National Science Foundation of China (Grants No. 11974311 and No. U1801661) and Zhejiang Provincial Natural Science Foundation of China under Grant No. LD18A040001.

\appendix

\section{Cooling coefficients}\label{DerivationCoefficient}

This appendix contributes to deducing the cooling coefficients for the measurement-induced cooling model in Eq.~(\ref{TotalHamiltonian}). In a general situation, a nonvanishing detuning $\Delta_e\equiv\omega_e-\omega_m\neq0$ exists between the resonator and the transition $|e\rangle\leftrightarrow|g\rangle$ in the ancillary system. In the rotating frame with respect to $\omega_m(b^\dagger b+|e\rangle\langle e|)+\omega_f|f\rangle\langle f|$, the total Hamiltonian becomes
\begin{equation}
{\tilde{H}'}_I=\Delta_e|e\rangle\langle e| +g_m\left(b^\dagger \sigma_{eg}^-+b\sigma_{eg}^+\right)+g_f\left(\sigma_{fe}^++\sigma_{fe}^-\right),
\end{equation}
under the RWA and the resonant-driving assumption, i.e., $\upsilon=\omega_f-\omega_e$.

\begin{figure}[htbp]
\centering
\includegraphics[width=0.4\textwidth]{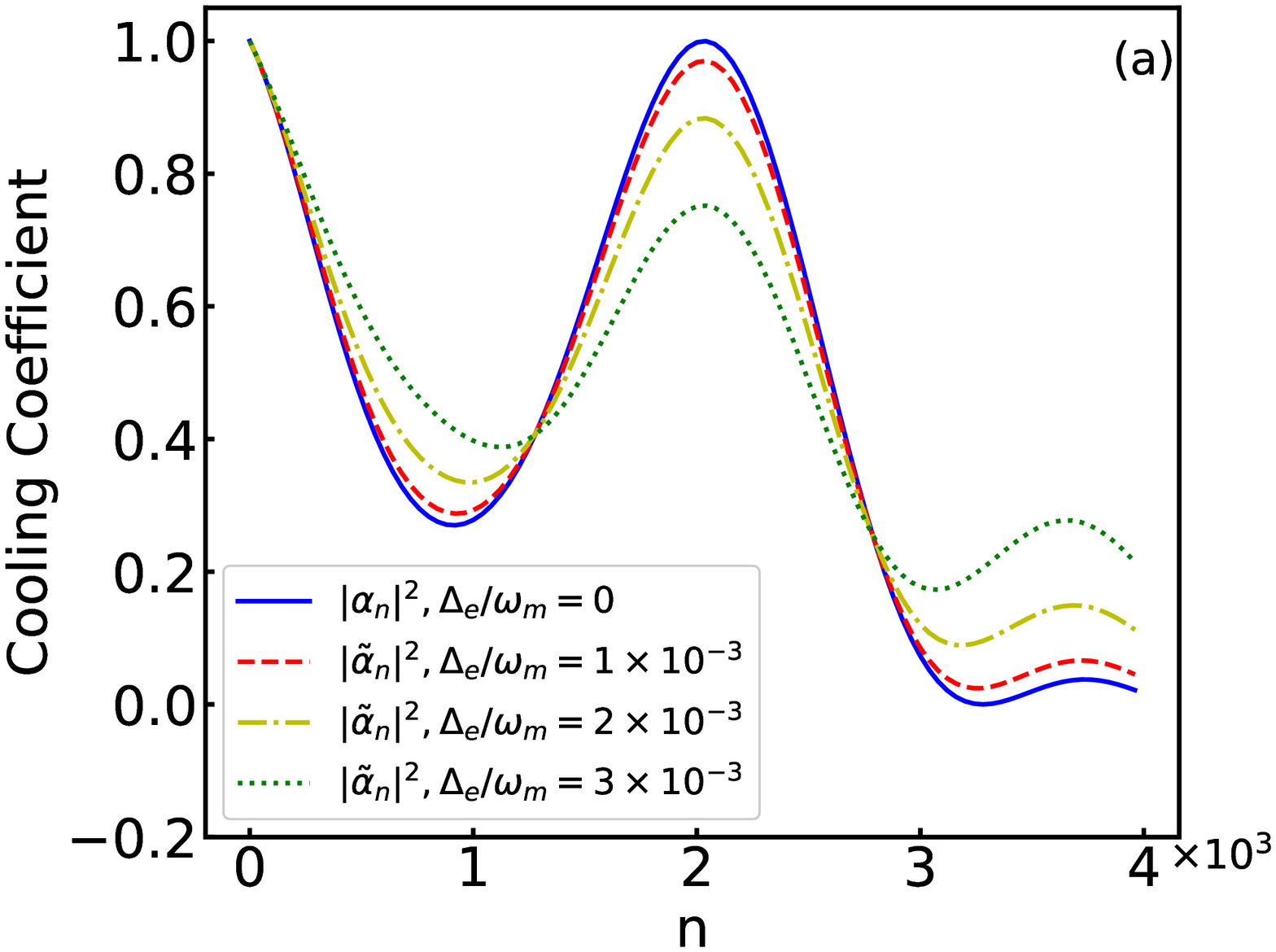}
\includegraphics[width=0.4\textwidth]{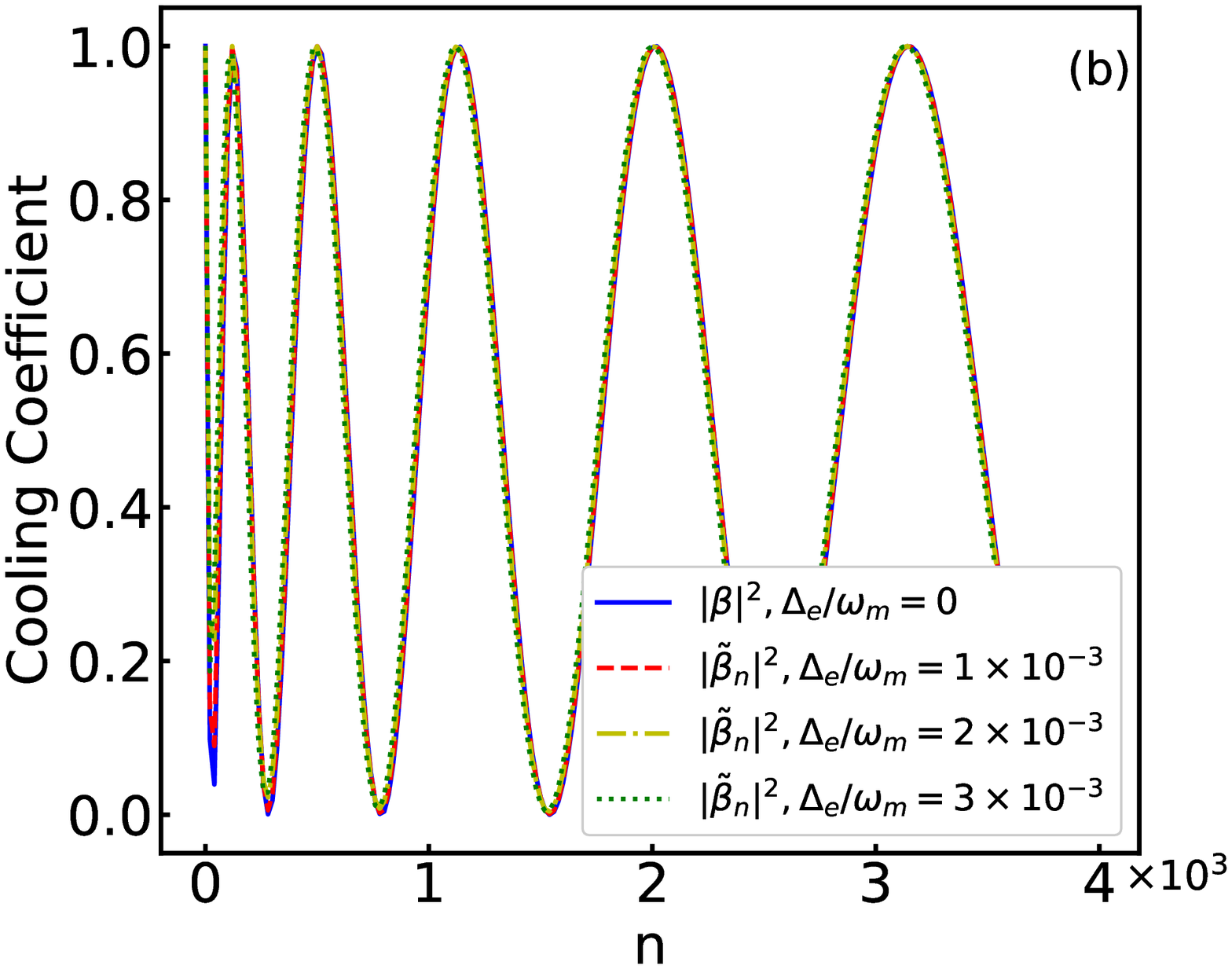}
\caption{Cooling coefficients of (a) the driving-assisted protocol $|\alpha_n|^2$ or $|\tilde{\alpha}_n|^2$ and (b) the conventional protocol $|\beta_n|^2$ or $|\tilde{\beta}_n|^2$ as functions of the Fock state index $n$ after a single measurement. $g_m/\omega_m=0.0004$, $g_f/g_m=50$, and $\omega_m\tau=700$.}
\label{CoefficientOffResonance}
\end{figure}

Similar to the JC Hamiltonian used in the conventional cooling method~\cite{MechanicalResonatorCooling}, ${\tilde{H}'}_I$ conserves the excitation number in the whole system. Then ${\tilde{H}'}_I$ is block-diagonal in the Hilbert space. It can be written as
\begin{equation}\label{HIn}
\tilde{H}_I^{'(n)}=\begin{pmatrix}
0&g_m\sqrt{n}&0\\
g_m\sqrt{n}&\Delta_e&g_f\\
0&g_f&0\end{pmatrix},
\end{equation}
in the $n$-excitation subspace spanned by $\{|g,n\rangle,|e,n-1\rangle,|f,n-1\rangle\}$. The free evolution operator during the measurement interval $\tau$ is straightforwardly written as $U(\tau)=\exp(-i{\tilde{H}'}_I\tau)$. The nonunitary evolution operator under a single measurement of the ground state [see the argument around Eqs.~(\ref{probability}) and (\ref{Vg})] is obtained by $V_g(\tau)\equiv\langle g|U(\tau)|g\rangle$, yielding the cooling coefficient
\begin{equation}
\begin{aligned}
&\tilde{\alpha}_n(\tau)=\langle n|V_g(\tau)|n\rangle
\\& =\frac{2g_f^2\tilde{\Omega}_n+e^{-i\frac{\Delta_e\tau}{2}}g_m^2n\left(i \Delta_e\sin{\tilde{\Omega}_n\tau}+2\tilde{\Omega}_n\cos{\tilde{\Omega}_n\tau}\right)}{2\Omega_n^2\tilde{\Omega}_n},
\end{aligned}
\end{equation}
where
\begin{equation}
\tilde{\Omega}_n\equiv\sqrt{g_m^2n+g_f^2+\frac{\Delta_e^2}{4}}, \quad \Omega_n\equiv\sqrt{g_m^2n+g_f^2}.
\end{equation}
Under the nondriving condition $g_f=0$, the whole model reduces to the JC model~\cite{MechanicalResonatorCooling}. Up to an irrelevant phase factor originating from various rotating frames, one can check that $\tilde{\alpha}_n(\tau)$ reduces exactly to
\begin{equation}
\tilde{\beta}_n=e^{-i\tau\Delta_e/2}\left(\cos{\Omega_c\tau}+i\sin{\Omega_c\tau}\cos{2\theta_n}\right),
\end{equation}
where
\begin{equation}
\Omega_c\equiv\sqrt{g_m^2n+\frac{\Delta_e^2}{4}}=\tilde{\Omega}_n|_{g_f=0}, \quad \tan{2\theta_n}=\frac{2g_m\sqrt{n}}{\Delta_e}.
\end{equation}

In the resonant case, i.e., $\Delta_e=0$, the rotated Hamiltonian in Eq.~(\ref{HIn}) becomes
\begin{equation}
{H'_I}^{(n)}=\begin{pmatrix}
0&g_m\sqrt{n}&0\\
g_m\sqrt{n}&0&g_f\\
0&g_f&0\end{pmatrix}.
\end{equation}
Consequently, the cooling coefficient becomes
\begin{equation}
\alpha_n(\tau)=\frac{\Omega_n^2+ng_m^2(\cos{\Omega_n\tau}-1)}{\Omega_n^2},
\end{equation}
as used in the main text. Under the same condition, the cooling efficient for the conventional protocol is
\begin{equation}
\beta_n(\tau)=\cos(g_m\sqrt{n}\tau).
\end{equation}

In Fig.~\ref{CoefficientOffResonance}, we plot the cooling coefficients for both protocols with various detunings. We can see in Fig.~\ref{CoefficientOffResonance}(b) that the quasiperiodic behavior of the coefficients $|\tilde{\beta}_n|$ for the conventional protocol is not sensitive to the detuning between the resonator and the excited state $|e\rangle$ of the ancillary system. In contrast, we can see in Fig.~\ref{CoefficientOffResonance}(a) that $|\tilde{\alpha}_n|$ for our protocol no longer behaves in a quasiperiodic manner. A larger $\Delta_e$ yields a wider cooling range, since $|\tilde{\alpha}_n|$ decays asymptotically with $n$. Thus, the cooling performance of our protocol is lower bounded by the resonant case with regard to the cooling range. Then we need to consider merely the resonant case in the main text.

\bibliographystyle{apsrevlong}
\bibliography{ref}

\end{document}